# Quantum-based vacuum metrology at NIST

Julia Scherschligt,[a] James A. Fedchak, Zeeshan Ahmed, Daniel S. Barker, Kevin Douglass, Stephen Eckel, Edward Hanson, Jay Hendricks, Nikolai Klimov, Thomas Purdy, Jacob Ricker, Robinjeet Singh, Jack Stone

National Institute of Standards and Technology, 100 Bureau Dr. Gaithersburg, MD 20899

[a] Electronic mail: julia.s@nist.gov

The measurement science in realizing and disseminating the unit for pressure in the International System of Units (SI), the pascal (Pa), has been the subject of much interest at the National Institute of Standards and Technology (NIST). Modern optical-based techniques for pascal metrology have been investigated, including multi-photon ionization and cavity ringdown spectroscopy. Work is ongoing to recast the pascal in terms of quantum properties and fundamental constants and in so doing, make vacuum metrology consistent with the global trend toward quantum-based metrology. NIST has ongoing projects that interrogate the index of refraction of a gas using an optical cavity for low vacuum, and count background particles in high vacuum to extreme high vacuum using trapped laser-cooled atoms.

## I. INTRODUCTION

As the national metrology institute of the United States, the National Institute of Standards and Technology (NIST) has responsibility to maintain and disseminate the unit of pressure, the pascal (Pa). Since its inception as the National Bureau of Standards (NBS) in 1901, NIST has advanced the science of pressure metrology, forging new techniques and technologies, as well as developing the science underpinning what it means to measure the pascal. Pressure metrology is particularly challenging in the vacuum, and especially in high vacuum ($<10^{-4}$ Pa) where the mean-free-path of molecules are longer than the dimensions of typical laboratory apparatus. Moreover, in the ultra-high vacuum (UHV, $<10^{-6}$ Pa)—a pressure regime critical to advanced research and technology[1] there has not existed an absolute pressure sensor. Recently, NIST has launched two initiatives to realize the pascal for vacuum pressures in a fundamentally modern way, by interrogations of quantum mechanical systems that directly relate to the particle density and therefore pressure in the vacuum. The *Fixed-Length Optical Cavity* (FLOC) is an index of refraction-based measurement. The *Cold Atom Vacuum Standard* (CAVS) uses cold trapped atoms to sense vacuum. These efforts are consistent with the nascent *Quantum-SI* which is an emerging effort in the international community to recast the SI (International System of units) in terms of observable quantum phenomena and fundamental constants of nature. Another nascent effort at NIST, the SiN ring-down membrane gauge, which we dub the "brane gauge", also has the prospective to be a Quantum-SI sensor in the vacuum. Past efforts at NIST have explored using resonant-enhanced multi-photon ionization (REMPI) and cavity ring-down spectroscopy (CRDS) techniques as tools for partial pressure analysis in the UHV and below, as well as spectroscopic techniques for measuring transient pressure.

Traditionally pressure is defined as a force per unit area, but as pressures extend further and further below an atmosphere (deeper into the vacuum) this definition becomes increasingly inconvenient and impractical. Instead, at low pressures the pascal is realized through the ideal gas law,

$$p = \rho_N k_B T = \rho_V RT , \qquad (1)$$

where $\rho_N$ is the number density of particles and $\rho_V$ is the molar density, $R$ is the gas constant, and $T$ is the temperature. In this formulation, pressure metrology becomes a counting problem, specifically, counting particles in the vacuum by any available technique. This reflects the applications as well: in the high-vacuum and below, most users are concerned with the amount of gas in the vacuum, e.g. as a contaminant, rather than the force it produces. Eq. (1) fundamentally relates pressure to the Boltzmann constant $k_B$, which will become a fixed constant with the redefinition of the SI in 2018.[2–4] With modern techniques and the trend away from artifact-based metrology, NIST and other institutes are developing the Quantum-SI, a metrology paradigm in which measurements are performed by making observations of quantum phenomena. With this new way of realizing the SI, the units are tied to defined physical constants, e.g. Plank's constant or the speed of light in vacuum. Furthermore, there is an accompanying shift away from electronic to photonic measurements. Measuring photons instead of electrons has several inherent benefits: optical signals are generally less to prone to pick-up noise from stray signals than are electrical signals, especially for long transmission distances. Photonic signals are high-fidelity, and can travel farther without regeneration. Additionally, optical fiber is lighter and has a larger bandwidth per cross-sectional area than copper wire, and can better handle harsh conditions, and so it has practical advantage, especially for use in aircraft or launch vehicles. Photonic measurements can be readily multiplexed and allow remote interrogation. Furthermore, photons can be used to directly probe the electronic states of atoms or molecules, and to prepare quantum states, making them the tool of choice for fundamental quantum measurements.

At pressures from about at atmosphere to the high vacuum, classical metrology technologies are mature and can deliver uncertainties at the level of a few parts in $10^6$, generally adequate for stakeholders. In consideration of this, the NIST efforts to recast the SI in terms of quantum effects should are not an attempt to further reduce uncertainties—though we hope that as the technologies occur this will become possible. Rather, by developing quantum-SI based techniques at these higher ranges, our goal is to enable stakeholders to have their own standards that are of the highest metrological integrity that never need calibration. Furthermore, these new technologies may enable the user to use the same device as a primary standard and a sensor, or as calibration-free sensors. Another advantage of pressure standard based on the FLOC technique is that it has the perspective to replace traditional mercury manometers, which are often used in the vacuum range of $10^{-3}$ Pa to $10^5$ Pa, thus removing toxic mercury from the calibration lab. The primary high-accuracy manometers used in this pressure range also tend to be rather large, expensive, and require a high level of expertise to operate, and are thus usually owned and operated by national metrology institutes or sophisticated calibration laboratories. The FLOC and the other quantum-SI techniques (such as the CAVS) presented in this review all have the perspective to be portable primary standards.

In the UHV and below, using photons to probe pressure is very appealing compared to the traditional ionization gauges and quadrupole mass spectrometers. These have been the subject of many reviews.[5–12] In these gas sensing techniques, ions are created via impact with electrons emitted from a hot-filament or, as is the case for a cold-cathode gauge, in a high-potential cathodic discharge. These ions are then detected by generating a current on an electrode or by an electron multiplier. Although these techniques have been the mainstay for UHV detection for several decades and many improvements have been made to make them more stable or to detect lower vacuum levels,[13–17] they have not been completely satisfying for measuring total or partial pressures in the UHV or extreme-high vacuum (XHV <$10^{-9}$ Pa) for several reasons.  First, the heat generated by these gauges cause sufficient outgassing to change the pressure in a vacuum system, second, the electron impact can "crack" molecules into fragments thus changing the chemical composition of the gas ( a particular problem in partial pressure analysis), third, the chemical composition can also be altered by chemical reactions on the hot filaments or other surfaces within the ionizer, fourth, the ionization technique does not produce a primary sensor, i.e., an absolute sensor that does not require calibration.  Additionally, electron-stimulated desorption (ESD) of ions from surfaces and the generation of X-rays due to electron impact on surfaces cause false signals. Photonic and quantum-SI methods have the potential to create absolute sensors without these problems. The heat-load on the vacuum system generated by photons is anticipated to be many orders of magnitude less than in ionization techniques. This reduces the possibility of changing the chemical composition of the gas and outgassing in the system. Most stakeholders for UHV or XHV metrology require uncertainties on the order of parts per hundred, but, as discussed above, presently there is no primary sensor in this vacuum range.

NIST has supported vacuum metrology through its calibration services and by developing and maintaining vacuum standards. Presently these cover the vacuum range down to $10^{-7}$ Pa.  These efforts support a wide variety of industries and research, such as semiconductor manufacturing, quantum information, particle physics facilities, space sciences, and nanotechnology. Developing quantum-SI standards to cover the present range of NIST's capabilities, as well as pushing standards to cover vacuum to $10^{-10}$ Pa or below (XHV), is a high priority. The goal is to create portable absolute sensors which are primary standards never requiring calibration, that can be owned by users outside of the national metrology laboratory. We are particularly motivated to develop quantum-SI sensors to cover the entire UHV range and below. NIST has a long and pioneering role in the field of ultra-cold atom physics. We visualize a new era of high metrological quality quantum-SI sensors based on cold atoms measure quantities such as time (which is already based on ultra-cold atoms), inertia, magnetic fields, gravity, and, of course, vacuum pressure. All such devices, and ultra-cold atom research in general, require UHV pressures or below to operate. Similarly, UHV quality is a concern in the field of quantum information. Building practical sensors and devices from cold atoms will require that UHV pressure be maintained over the lifetime of the device. One suggested metric for this is $10^{-8}$ Pa for 1000 days.[18] The vacuum requirements are a technical challenge in creating such devices. We are presently developing a portable metrology device for deep vacuum that is simultaneously a quantum-SI standard and a sensor. The Portable, Intrinsic, Cold-atom, Optical Vacuum Standard or PICO-VS which will not only address these technical challenges, but will then be a tool for quantum research and development.

In this paper, we begin with a brief overview of traditional vacuum metrology, then discuss early work to move beyond artifact-based measurements: multiphoton ionization and cavity ringdown spectroscopy. We then cover in more detail the two major optical pressure projects presently underway at NIST to develop new realizations of the pascal consistent with the emergent Quantum-SI paradigm: The Fixed-Length Optical Cavity (FLOC) which operates at pressures from 1 Pa to ≳ 100 kPa, and the Cold Atom Vacuum Standard (CAVS) which operates from ultra-high vacuum to extreme-high vacuum (UHV to XHV, or from about $10^{-6}$ Pa to < $10^{-9}$ Pa). We discuss how this new approach will enable the next generation of practical, deployable sensor technologies. Finally, we will describe a new research effort to develop the brane-gauge, followed by a description of NIST's work on spectroscopic techniques for measuring transient pressure. Special attention will be paid to the lower pressure limits anticipated in these new standards and sensors.

## II. TRADITIONAL PASCAL

The concept of metrology coevolved with the that of commerce as early as 3100 BC in Mesopotamia, and was the precursor to the development of both western mathematics and written language.[19] For thousands of years until the last century, the science of measurement relied entirely upon comparisons between objects of interest and standard artifacts, but since the advent of modern physics, new ways to realize units of measure have begun to take hold that are based on immutable properties of nature, particularly for length (based on the speed of light) and time (based on quantum properties of atoms). Pressure is traditionally defined as force per unit area, $P = F / A$, and has units of pascal (1 Pa = 1 N m$^{-2}$). Therefore, to generate or realize the pascal, the most obvious method is to apply a known force to a known area. This is the operating principle behind the piston gauge, the workhorse primary pressure standard for pressures around an atmosphere (100 kPa) to a few hundred megapascal. Piston gauges consist of a piston and cylinder assembly with well-characterized dimensions—for proper primary standards, the area of the piston gauge is measured using primary dimensional metrology and corrected for distortion effects with careful numerical modeling. The gauge is then loaded with mass units that have been independently characterized using standard techniques in mass metrology. The combination of known mass and known area gives pressure. Though the details of operation have been modernized and refined, the underlying concept of the technique is ancient.[20,21]

For measurements at atmospheric pressures and into the low vacuum, manometry is the traditional technique. The manometer is generally considered to be invented by Torricelli in the seventeenth century,[22] and though it has been incrementally refined and improved over the centuries, it has remained the state-of-the-art until now. Manometers operate on the principle that a fluid in a column sealed at the top will create a vacuum in the sealed end of the column when it experiences the downward force due to its own weight. The pressure on the other end of the column (the pressure of interest, often atmosphere) exerts a force that must balance the gravitational force, for the fluid to be in equilibrium. The pressure in pascal is then $P = \rho_f g h$ where $\rho_f$ is the fluid density, $g$ is the local acceleration due to gravity, and $h$ is the column height. NIST operates Ultrasonic Interferometer Manometers (UIMs), with mercury as the fluid (with a full scale range of 360 kPa) and with oil as the fluid (with a range of 0.1 Pa to 120 Pa). The determination of column height is done using an ultrasonic technique, and care is taken to minimize uncertainty from other sources including temperature. These instruments can claim relative standard uncertainties as low as 3 x 10$^{-6}$ as demonstrated in an international key comparison.[23,24]

At lower pressures, it becomes much more convenient to formulate the pascal as the translational kinetic energy density of particles in a volume (1 Pa = 1 J m$^{-3}$), rather than a force applied to an area as defined above. To generate pressures in the high vacuum and ultra-high vacuum, a commonly used method is to use a flowmeter with dynamic expansion technique. In this technique, a known flow of gas $\dot{n}$ is injected into a vacuum chamber upstream of a flow constrictor with a known conductance $C$. In the molecular flow regime (were the mean-free path is larger than the vacuum vessel or flow constrictor), the pressure difference across the flow limiter is given by the pressure analogy to Ohm's law

$$p_{\text{upper}} - p_{\text{lower}} = \dot{n}RT/C, \qquad (2)$$

which tells us that the pressure difference across an orifice is the flow divided by the conductance. The upstream pressure, $p_{\text{upper}}$ is the pressure above the flow constrictor (typically an upper chamber in a vacuum system), and $p_{\text{lower}}$ is the pressure downstream of the flow constrictor (typically a lower chamber in a vacuum system). A high pumping-speed vacuum pump is connected to the lower chamber such that $p_{\text{upper}} > p_{\text{lower}}$. If the ratio $p_{\text{upper}} / p_{\text{lower}}$ is known from a separate measurement, or $p_{\text{upper}} \gg p_{\text{lower}}$ and $p_{\text{lower}}$ can be neglected, then a standard pressure $p_{\text{upper}}$ may be determined from the known C and $\dot{n}$. To produce a known flow of gas $\dot{n}$ with low uncertainty, a constant pressure flowmeter may be employed whereby a known flow, $\dot{n}$, from a leak in a volume $V(t)$ is determined by inducing a volume change $\dot{V}$ to hold the pressure $p$ within the volume constant.[25–27] We see from Eq. (1) that the gas flow can be written $\dot{n} = p\dot{V}/N_A k_B T$, where $N_A$ is Avogadro's constant and the gas flow $\dot{n}$ has units of mol/s. The flowmeter plus dynamic expansion apparatus together constitutes the present state-of-the-art standard for high vacuum and ultra-high vacuum. However, it should be noted that this system fails to meet the technical definition of primary (for pressure) according to the International Vocabulary of Metrology (VIM) because the flowmeter relies on a calibrated pressure gauge.[28] Still, it is *functionally* primary[29]—many national metrology institutes (NMIs) which operate these standards calibrate these pressure gauges using primary methods—, and is used extensively at NIST, the metrology institute of Germany (Physikalisch-Technische Bundesanstalt or PTB), and other NMIs for calibrations of vacuum gauges, notably ionization gauges and spinning rotor gauges.[25,31–35]

In 2018, the year this article is published, the values of physical constants will be fixed by the International Committee for Weights and Measures (CIPM) with profound consequences on metrology in general and pressure metrology in particular. With fixed values of the Boltzmann constant and Avogadro's number, direct measurement of $\rho_N$ or $\rho_V$ gives pressure absolutely (assuming that temperature uncertainty can be suppressed sufficiently low as to be negligible.) In this paper, we emphasize two methods under development for assessing number density $\rho$, which become *de facto* primary pressure standards upon redefinition of the SI. The first is to measure the refractive index of the gas at the pressure of interest, the second is to measure the lifetime of a trap of cold atoms bombarded by gas molecules in the volume. Before discussing these current projects, we turn to pioneering efforts at NIST to measure vacuum using photonic-based spectroscopic techniques.

# III. OPTICAL METHODS FOR MEASURING PARTIAL PRESSURES AT NIST
## A. *Resonant-enhanced multi-photon ionization*

Multi-photon ionization (MPI) can be used to ionize molecules which can be subsequently detected using traditional techniques such as by electron multipliers or multichannel plates. It has advantages over electron-impact ionization techniques based on, for example, hot filaments, which tend to outgas, promote chemical reactions, and produce indiscriminate fragmentation of gases. In the 1990's, Looney and coworkers made quantitative partial pressure measurements of CO using the laser-based technique of resonant-enhanced multi-photon ionization (REMPI) techniques.[36,37] They found it possible to detect CO partial pressures as small as $10^{-10}$ Pa, and demonstrated the ability to measure partial pressures of $10^{-9}$ Pa with an uncertainty of 20 to 30%. In REMPI, a molecule is excited by one or more photons to an electronic intermediate state, and subsequently ionized by absorbing one or more photons from the intermediate excited state. CO is ionized via a three-photon process: a two-photon excitation using 230 nm laser light promotes the molecule from a $X\,^1\Sigma^+$ state to the $B\,^1\Sigma^+$ state, where the molecule is subsequently photo-ionized by another 230 nm photon. Resonant ionization techniques have the advantage over non-resonant techniques in that it is selective in gas species, making it very sensitive detection technique for specific gases. Previous to the NIST work, REMPI had already shown promise as a sensitive detection technique,[38] and continues to be an active field today. The work done at NIST by Looney and coworkers demonstrated the first quantitative REMPI measurements. They used a time-of-flight (TOF) mass spectrometer to detect ionized CO. The TOF spectrometer was capable of resolving CO from $N_2$, but no ionized $N_2$ was detected, thus demonstrating the excellent species selectivity of the REMPI technique. The TOF mass spectrometer signal was calibrated against a spinning rotor gauge using a split-flow technique, thus enabling quantitative partial pressure detection of CO. The REMPI technique is an excellent way to detect specific gases in the UHV and XHV, and is particularly useful for chemically active gases. In order for the REMPI technique to be used for absolute measurements of partial pressure, the accurate cross-sections for multi-photon ionization must first be determined, which is a difficult task and remains outstanding for many molecules.

## B.  *Cavity ring-down spectroscopy*

Another highly sensitive optical detection technique is that of cavity ring-down spectroscopy (CRDS) shown schematically in Figure 1.  NIST began a program to develop CRDS into a highly sensitive quantitative tool for the detection of molecules the 1990s.[39,40] A laser pulse is injected into a high-finesse optical cavity defined by two highly reflective mirrors of reflectivity $R$ separated by the cavity length $l$. The output intensity will have a "ring-down" time given by the expression[41]

$$\tau(\omega) = \frac{l}{c\left[(1-R) + \alpha(\omega)l\right]}, \qquad (3)$$

where $\alpha(\omega)$ is the absorption coefficient of the gas within the optical cavity.  $\alpha(\omega)$ can be determined from the difference between the cavity ring-down time for an empty cavity and that containing the gas of interest.  In fact, the CRDS technique is a powerful tool for measuring absorption coefficients,[42] particularly those for weak transitions. The absorption coefficient is related to the number density of the gas $\rho_N$, the line strength of the absorption transition $S$, and the normalized line-shape function $f(\omega)$:

$$\alpha(\omega) = 2\pi c \rho_N S f(\omega). \qquad (4)$$

Thus if $S$ and $f(\omega)$ are known, the number density $\rho_N$ and hence gas pressure can be determined from the ring-down time.

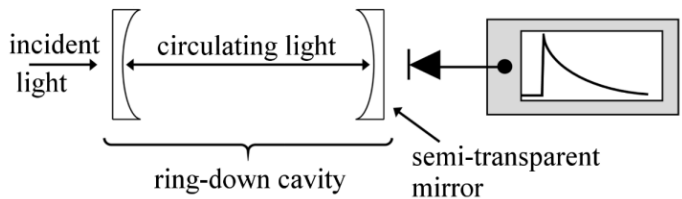

Fig. 1. A schematic diagram of a cavity ring-down spectroscopy apparatus. Reproduced from R. D. van Zee, P .J. Looney, and J.T. Hodges, in Proc. SPIE Adv. Sensors Monit. Process Ind. Environ. (1999), pp. 46–56.

Like the REMPI technique, the CRDS method is most useful for sensing specific gases. In principle, it can be used to sense virtually any molecule, with the practical caveat that the molecule must have an absorption transition which is both strong enough to do CRDS, and whose energy corresponds to wavelength accessible by available lasers. The CRDS method has been shown to be capable of sensing $CO_2$ concentrations at the level of 43 parts in $10^{15}$.[44] As pointed out in Jousten *et al*.,[45] this corresponds to a partial pressure of $4.3 \times 10^{-9}$ Pa; however, it is not clear that the CRDS method can be used to detect UHV or XHV partial pressures for an arbitrary gas. As discussed in van Zee *et al*.,[43] there is a minimum detectable absorptivity which depends inversely on $\tau^2$ and inversely on the square root of the number of measurements. This means that UHV or XHV measurements require a minimum absorption strength $\alpha$ for a given cavity length and data acquisition time. From the examples given in van Zee et al. (see Fig. 2), using CRDS to detect UHV partial pressures for molecules like CO or $CO_2$ may be possible, but it may not be practical for molecules such as $H_2O$ or $C_2H_2$. For partial pressure measurement, much of the NIST program has focused on detecting concentrations of gas in nominally atmospheric pressures, such as $O_2$ or $H_2O$ in $N_2$. The NIST program has been successful in performing highly accurate measurements of water vapor pressure;[46,47]

molar fractions of water vapor equal to $7\times10^{-8}$ have been determined.[48]

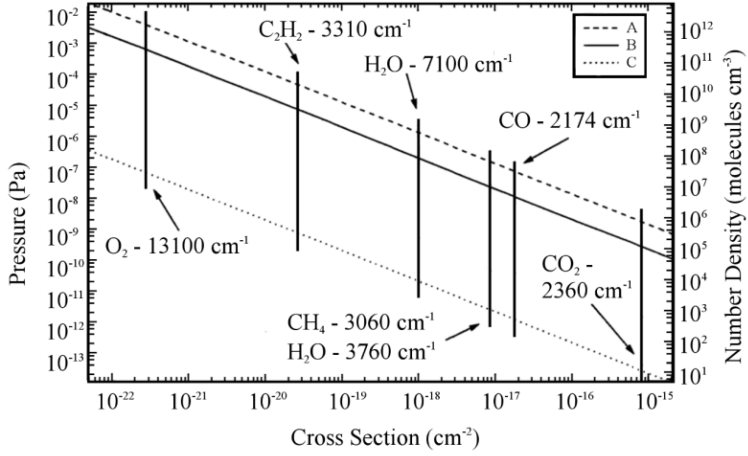

FIG. 2. A plot of the lowest number density measurable during a one second measurement interval as a function of cross section for three sensitivities: A – as demonstrated in van Zee et al.[43] (10 cm cavity with a mirror reflectivity of 0.999996, B—shot noise limit for these experiments; C—shot noise limit for a1 m long cavity mirror reflectivity of 0.99999, and 100 μW of laser power exiting the cavity. Reprinted from R. D. van Zee, P .J. Looney, and J.T. Hodges, in Proc. SPIE Adv. Sensors Monit. Process Ind. Environ. (1999), pp. 46–56.

## IV. THE QUANTUM PASCAL: OPTICAL REFRACTOMETRY

### A. *Underlying principle*

We now turn our attention to the first of our active research projects in vacuum metrology at NIST. Several laser-based interferometer techniques are under study to interrogate the refractivity $n-1$ of a gas ($n$ is index of refraction) which is a proxy for the gas density $\rho_N$, and ultimately the pressure $p$ through the equation of state:[49]

$$p = k_B T \rho_N (1 + B_\rho \rho_N + C_\rho \rho_N^2 + ...), \tag{5}$$

where $k_B$ is the Boltzmann constant, $T$ is thermodynamic temperature, and the deviations from the ideal gas law arising from two- and three-body interactions are taken into account by density virial coefficients $B_\rho$ and $C_\rho$. For helium gas, the virial coefficients in

(5) are calculable through statistical mechanics at a level that contributes less than one part in $2 \times 10^7$ to the uncertainty of pressures below 1 MPa.[50] Current state-of-the-art thermodynamic thermometry implies that the thermal energy $k_B T$ can be measured better than one part in $10^6$.[51] Therefore, with the highest accuracy measurements of helium refractivity, uncertainties from theory and thermodynamic temperature imply that the pascal can be realized with uncertainty at the one part in $10^6$ level, which would place it competitive with state-of-the-art piston gauges at 1 MPa, and better than state-of-the-art mercury manometers at 100 kPa and below.

Depending on the details of these approaches, the techniques described herein result in a device that is considered alternately functionally-primary, primary, or a transfer standard. In all cases, two major obstacles must be overcome which are discussed below: The pressure-dependent index of refraction must be known to high accuracy, and any distortions in the measurement device must be accounted for. We begin with a brief discussion of the underlying physics before turning to a description of several experimental devices. The speed of light with frequency $v$ in a gas, $c$, is reduced from that in an ideal vacuum $c_0$ by a coefficient $n$, that is,

$$c = c_0/n . \qquad (6)$$

The mechanism by which this happens concerns the polarizability of the particles constituting the gas. Such polarizabilities are the quantum basis of the method, and our ability to calculate the polarizability of helium and thus its refractivity is ultimately what makes the technique described herein a fundamental standard, consistent with the quantum-SI. Theoretic determinations of these fundamental atomic properties were performed at relativistic and quantum electrodynamics (QED) levels.[52] Extending the method to gases other than helium is done in a ratiometric way that preserves the fundamental nature of the method.

The relation of $n$ to $\rho_N$ for an isotropic homogeneous medium is obtained by the Lorentz-Lorenz equation,[53]

$$\frac{n^2-1}{n^2+2} = \frac{1}{3\varepsilon_0}\rho_N\alpha = A_R\rho_V, \qquad (7)$$

where $\alpha$ is the dynamic polarizability of an individual molecule of gas in the volume, $A_R$ is a virial coefficient, the molar dynamic polarizability, and $\varepsilon_0$ is a fixed physical constant, the vacuum dielectric permittivity. Thus by determining index refraction, we can realize $\rho_V$. To calculate polarizability from first principles requires taking into account relativistic, QED, and finite mass effects[52] and this has been done for both the polarizability and refractive index of helium to an uncertainty of below one part in $10^6$. (note that for accuracy on the order of one part in $10^6$, it is also necessary to include the effect of magnetic susceptibility, which is omitted in Eq. (7) for simplicity.

Pressure sensors based on refractometry can in principle be based on any gas and He has the advantage that it's pressure dependent index of refraction has been calculated to high accuracy, making such a device intrinsically absolute. However, in a practical device made of ultra-low expansion (ULE) glass, helium has the disadvantage that it is absorbed into the glass.[54] And so a refractometer using gases other than helium, such as $N_2$, may be a more useful method of pascal dissemination, but first the index of refraction of that measurement gas must be determined.

## B.  *Refractometers as pressure standards*

In this section, we will discuss how refractometers have been demonstrated to serve as pressure standards before finally discussing them as *primary* pressure standards in section C. The concept of index of refraction is that a photon with a fixed wavelength will have a different frequency in the presence of gas than in a vacuum as described in Pendrill.[55] This suggests an experiment in which one directs a laser down each of two channels, one filled with gas and the other evacuated, and measures the frequency change. This is done in the NIST *Fixed-Length Optical Cavity* (FLOC). More precisely, a laser is wavelength-locked in resonance to a Fabry-Perot cavity, if gas density (i.e. pressure) changes, the servo adjusts the frequency $f$ to maintain resonance with the cavity. Changes in $f$ then give the index of refraction according to:

$$n-1 \approx \frac{-\Delta f + \Delta m \left( c_0 / 2L \right)}{f} \; , \qquad (8)$$

where $\Delta f = f - f_0$ ($f_0$ is the laser frequency in vacuum, and $f$ is the frequency in the gas medium,) $\Delta m$ is the change in mode order, and $L$ is the length of the cavity. In practice, the laser frequency in eqn. (8) is never measured directly but is determined by measuring the difference in frequency between the measurement laser and a reference laser locked to the vacuum channel. Both the reference and vacuum channel deform under pressure. Much of the deformation is an overall compression due to finite bulk modulus, which is common to both the reference and measurement channels so that the effect largely cancels out. Another important effect is bending of the mirror surfaces in the reference channel due to the pressure differential across these mirrors. The measurement equation for pressure determined by the FLOC is then:[56]

$$p = \frac{1}{\left(\dfrac{3}{2k_BT}\right)A_R - d_m - d_r}\left(\frac{f_{vac} - f_{gas}}{f_{gas}}\right), \tag{9}$$

where $f_{vac}$ ($f_{gas}$) is the frequency in the evacuated (gas-filled) cavity. The distortion term $d_r$ is essentially the fractional change in length of the reference cavity when gas is added to the cavity (a negative number). Similarly, $d_m$ is the negative of the fractional change in the measurement cavity length (a positive number, where the sign is an artifact of the derivation). For simplicity, in Eq. (9) we have only retained terms of order $\Delta f/f$. The correction for the distortion terms are approximately $d_m \approx -d_r \approx 1.1 \times 10^{-11}$ Pa$^{-1}$, whereas the index $n$ varies with $p$ by $3.2 \times 10^{-10}$ Pa$^{-1}$ for helium at 303 K. Note that the two correction factors cancel each other within 10%. Therefore, without any correction for the distortion, the FLOC is a primary standard for pressure to about 0.3 %.

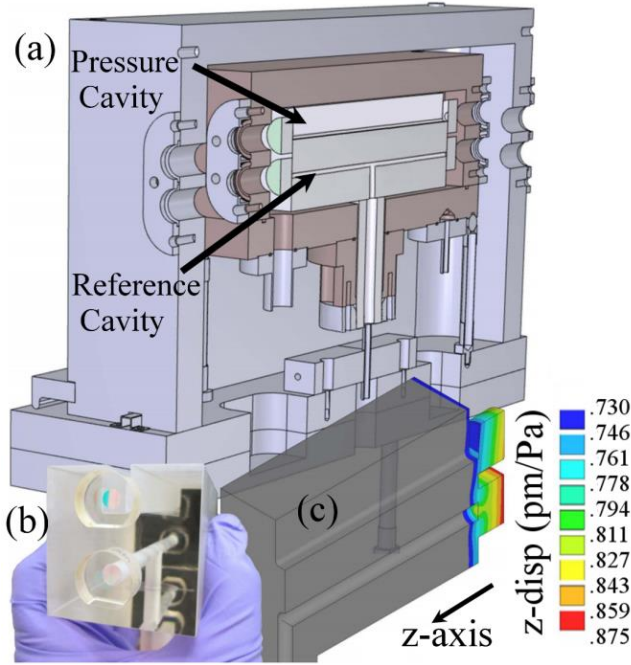

FIG. 3. (Color Online) (a) Dual FP cavity refractometer in its thermal/vacuum apparatus: the pressure measurement cavity is in gas, and the reference cavity is ion-pumped to high-vacuum. (b) photograph of the refractometer. (c) Distortions in cavity lengths per pascal of pressure on the measurement cavity when the reference cavity is at vacuum. Reproduced from P. Egan, J. Stone, J. Hendricks, J. Ricker, G. Scace, G. Strouse, Optics Letters 40, No. 17, 3945 (2015).

Much improved performance can be achieved by measuring two or more different gases of known refractivity at a certain pressure. Both the cavity distortion and the absolute pressure can be determined, since measurements of two gasses provide two equations in the two unknowns. Helium refractivity is known as a function of pressure by calculation; at present, nitrogen refractivity has been measured.[57] When a measurement is made using two gasses, the FLOC provides traceability to primary methods and becomes functionally primary in the important sense that it never needs to be calibrated against a pressure standard. Thus, the invariant atomic/molecular properties of the gasses (i.e., refractivity) will serve as a practical functional standard for universal dissemination of the Pascal. In past work, the FLOC demonstrated $((2 \text{ mPa})^2 + (8.8 \times 10^{-6} \, p)^2)^{1/2}$ expanded uncertainty as a transfer of the pascal, and so the FLOC as a transfer standard of the pascal outperforms the manometer at pressure below about 1 kPa.[56]

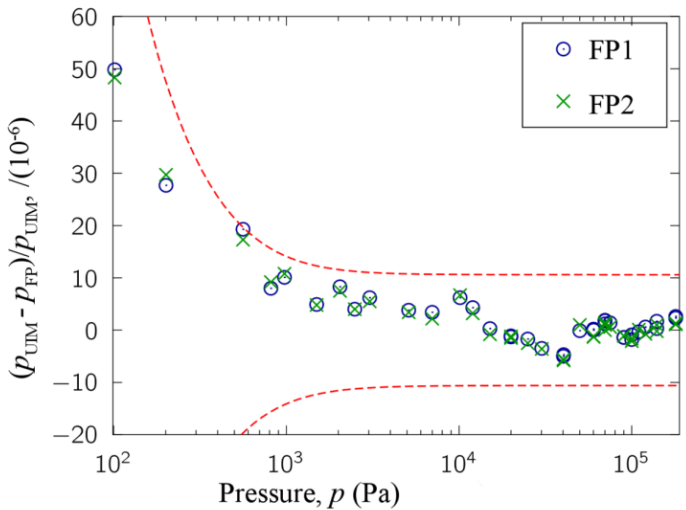

FIG. 4. (Color Online) Disagreement in pressure as measured by two separate laser refractometers ($p_{FP}$) and mercury ultrasonic manometer ($p_{UIM}$). The dashed lines are the manometer uncertainty. The figure is reproduced from P. F. Egan, J. A. Stone, J. E. Ricker, and J. H. Hendricks, Rev. Sci. Instrum. 87, (2016).

## C.  Methods for casting refractometers as primary standards

As described earlier, the FLOC is already a primary pressure standard when used with helium gas, but distortion of the optical cavity and mirrors, including dynamic effects caused by diffusion of helium into the ULE glass, limits the uncertainty to a level that is too high for many applications. Even if the measurement gas is nitrogen or some other species that doesn't diffuse into the glass, distortion still needs to be accounted for. What this means from a practical standpoint, is that to use a refractometer as a primary standard, we need to perform an excellent characterization of the distortion. At present, correcting for the distortion error in the FLOC device is being pursued by several different methods that are not first-order dependent on a measurement of pressure. These are outlined in turn in this section.

One early effort is shown in Fig. 5, in which an optical technique is employed to find the laser beam location on the mirror surface, and the shape is calculated through a finite-element analysis. From this, a bending profile is extracted. By combining the bending profile with knowledge of the beam location, an estimate can be made of the distortion error in the FLOC.  We have performed this procedure on two separate FLOC devices, and found agreement within a relative uncertainty of $7 \times 10^{-5}$ when compared with distortion determined by two-gas measurements (previous section).  The approach appeared equally limited by how accurately the geometry and beam location can be determined by the described imaging technique, and the 2 % uncertainty in the elastic properties of ULE.

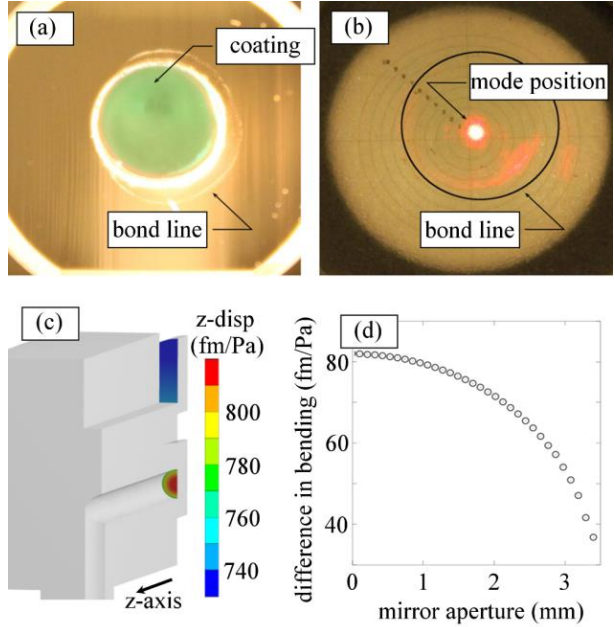

FIG. 5. (Color Online) Correcting FLOC distortion via finite-element analysis and an inspection of the mode position on the mirror. Pane (a) is an image of the mirror showing the bond interface. Through edge-detection, an estimate can be made of the area upon which the pressure acts. In (b), another image is taken with a laser beam aligned to the cavity resonance. By combining these two images, an estimate of the location of the beam on the mirror surface is made. The result of a finite-element analysis is shown in (c) datasheet values were used for elastic properties of ULE glass, and the geometry was estimated by the bond line in image (a). The difference in mirror bending calculated by finite-element is extracted as a profile, shown in pane (d).

Another possibility would be to determine the elastic properties of the glass directly by mechanical means, using resonant ultrasound spectroscopy like that described in Schmidt et al.[58] Achieving relative uncertainty lower than one part in $10^5$ in helium refractivity would require determination of the bulk modulus within 0.03 %, which to our knowledge has not previously been demonstrated with glass. Additionally, doping inhomogeneities in ULE (i.e., giving rise to variations in the coefficient of thermal expansion) are a concern, in the sense that a token whose elastic properties are measured by mechanical means may not be an accurate reflection of the elastic properties of the FP cavity itself. (High-purity fused silica may be the better choice; our experience with sub-milikelvin thermal stabilization suggests that the higher thermal expansion of fused silica will not adversely affect low pressure performance.)

A further possibility is to use multi-wavelength interferometry and calculated dispersion of helium to determine the FLOC distortion. This can be accomplished by interrogating the FLOC with two laser frequencies $v_1$ and $v_2$ locked to the optical cavity, which has the advantage that it can be done *in-situ*. The measurement equation for pressure determined by the FLOC under these conditions to first order is:

$$p = \frac{1}{\delta\alpha} \frac{k_B T}{2\pi} \left[ \left(\frac{\delta v_1}{v_1}\right) - \left(\frac{\delta v_2}{v_2}\right) \right] . \qquad (10)$$

Here $\delta\alpha$ is the change in the atomic polarizability between the two laser frequencies at the same gas pressure $p$. Again, the atomic polarizability $\alpha(\lambda)$, where $\lambda=c/v$, is known for He from fundamental theoretical calculations. We see in Eq. (10) that the distortion terms that were present in Eq. (9) have cancelled and thus, using two lasers, we now have a primary FLOC. The main disadvantage to the two-laser method is that dispersion is a small effect compared to refractivity. For two practical laser frequencies, say 633 nm (HeNe laser) and 1550 nm (standard telecom wavelength), the difference in *n*-1 is approximately $1.6 \times 10^{-7}$ (at atmospheric pressure), which is more than 200 times smaller than the value of $n - 1$. Some sources of noise and systematic uncertainties will increase, and the current state of theory and calculation of helium dispersion would limit the approach to 5 parts in $10^6$. Efforts are presently underway at NIST to create a primary FLOC using this multi-color technique.

The last approach we discuss to solving the distortion problem, and thus making the refractometry technique fully primary, is perhaps the most obvious. The distortions which currently limit FLOC performance as a primary pressure standard can be avoided and/or corrected in refractometers of alternate design. One such design is the Monolithic Interferometer for REfractometry (MIRE).[59] One key feature of the apparatus is three interchangeable triple-cells of different length as shown in Fig. (6), but almost identical geometries, material properties, and location of the laser beams through all windows. This feature is designed to make the window distortion common-mode in measurements of helium refractivity performed in cells of different lengths, and allowed us to cancel the error to 1.8 %, which resulted in a 9.8 ppm relative uncertainty in the refractometer. When the uncertainty in the refractometer was combined with the uncertainties in the thermodynamic temperature of helium, gas purity, and the Boltzmann constant, our total standard uncertainty in this primary realization of the pascal was 11.7 ppm.

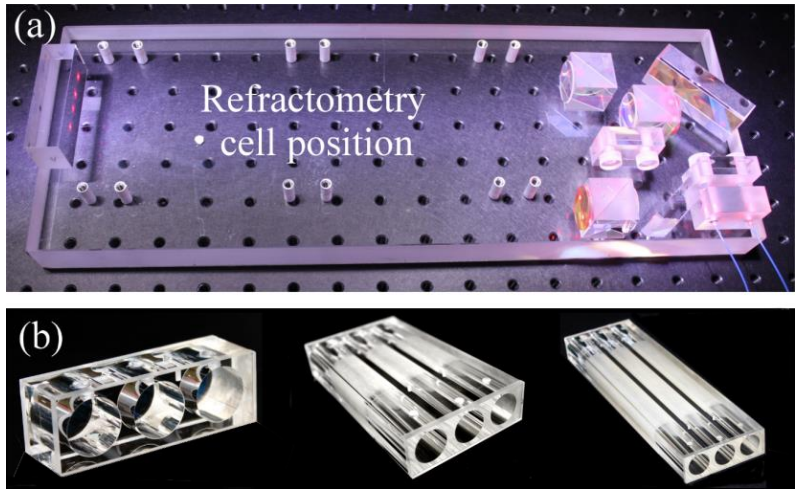

FIG. 6. (Color Online) (a) MIRE apparatus and (b) Refractometry cells of three different lengths but which are otherwise nominally identical. Each borehole has a gas inlet and outlet. (Left to right, the cell lengths are 18 mm, 134 mm, 254 mm.) Figure is based off of P. Egan, J. Stone, J. Hendricks, J. Ricker, G. Strouse, Optics Letters 42, No. 15, 2944 (2017).

Another alternative refractometer design is a variable length optical cavity (VLOC).[60] The VLOC differs from the FLOC in that it measures a pathlength $n \times \Delta L$ instead of $\Delta n \times L$; which is to say, that one changes the geometric length of a FP cavity filled with gas at a constant refractive index of helium by about 15 cm instead of changing the refractive index inside a FP cavity of 15 cm constant length. The concept of translating a mirror to avoid pressure-induced distortion is not necessarily immune from spring-induced distortions, and the complications of motion and geometry errors are an engineering challenge. The VLOC device is currently under development.

# V. QUANTUM PASCAL: COLD-ATOM VACUUM STANDARD

NIST is developing new a method for measuring and understanding the pascal at the lowest pressures, the Cold-Atom Vacuum Standard (CAVS) which uses a cold atom trap to sense pressure.[61] This work began in earnest in 2016. Since the earliest days of neutral atom trapping, it has been known that the background gas in the vacuum limits the trap lifetime (the characteristic time that atoms remain trapped). We are inverting this problem to create a quantum-based standard and sensor. Because the measured loss-rate of ultra-cold atoms from the trap depends on a fundamental atomic property (the loss-rate coefficient, related to the thermalized cross section) such atoms can be used as an *absolute* sensor and primary vacuum standard. Researchers have often observed that the relationship between the trap lifetime and background gas can be an indication of the vacuum level, and several research groups have pursued using cold atom traps as vacuum sensors.[62–67] However, an absolute vacuum standard, sufficient for use as an international quality standard, has not yet been realized. To do this requires rigorous attention to all potential error sources, from both the atomic perspective and the vacuum perspective. Moreover, a primary CAVS requires the collision cross section between trapped ultra-cold atoms and the background gas, discussed below, to be traceable to an *ab initio* theoretical determination. This work is ongoing at NIST, and much progress has already been made. In this section, we describe the operating principle of the CAVS, we discuss the sources of error and our approach to minimizing quantify said error, and present some initial data to illustrate the device's ultimate potential.

## A. Basis of the technique

The operational premise of the CAVS is that an individual atom is knocked out of the trap when it undergoes a collision with a background gas molecule, and that measuring the trap lifetime is therefore a way to count background particles. This naïve picture does a surprisingly good job of approximating the real behavior of the CAVS under the right conditions. That is, if we suppress other loss channels for trapped atoms so that we have a one to one correspondence between collision events and ejections, and if we have a good understanding of the collision probability so that we can rigorously relate the number of collisions to the number of background particles, then indeed we have a fundamental measure of background gas density. We address these two issues in turn, beginning with the latter.

## B. Collision cross sections

The trap loss mechanism of interest, that due to collisions between sensor atoms and background molecules, follows an exponential form. (Other loss channels may have different functional form, but for the moment we neglect these.) Then the number of atoms in the trap $N$ has the following time dependence,

$$N(t) = N_0 e^{-\Gamma t} . \tag{11}$$

The loss rate of the trap $K_{\text{loss}}$ is defined as the thermalized average of the collisional cross section $\sigma$ times the velocity $v$ of a particle, so that $K_{\text{loss}} = \langle \sigma v \rangle$. Therefore the decay rate $\Gamma$ is

$$\Gamma = \rho K_{\text{loss}} . \tag{12}$$

The background gas density $\rho$ is what we're ultimately trying to determine, so our goal is to make a measurement of $\Gamma$ and combine it with a theoretically calculated $K_{\text{loss}}$. Fully *ab initio* calculations are currently underway at NIST for the lithium plus molecular hydrogen system. This is a tractable five-electron system, and we anticipate the uncertainty associated with these calculations to contribute at the five percent level or less. Semiclassical estimates of $K_{\text{loss}}$ based on published $C_6$ or Casimir-Polder potentials have been carried out for a variety of systems, not just Li + $H_2$, but other sensor atom species such as Na, K, Rb, and Cs, with a variety of background gas species such as He, Ar, $N_2$, $O_2$, $H_2O$, $CO_2$ and others.[61,68] The overall trend shows little variation as a function of background gas species for all species other than hydrogen, with $K_{\text{loss}}$ varying from about $2 \times 10^{-9}$ cm$^3$s$^{-1}$ to $3 \times 10^{-9}$ cm$^3$s$^{-1}$, for hydrogen, the semiclassical estimate gives $K_{\text{loss}} \simeq 5$ cm$^3$s$^{-1}$. The variation of $K_{\text{loss}}$ with alkali sensor atom increases with sensor atom mass, i.e. $K_{\text{loss,Cs}} \simeq 1.5 \times K_{\text{loss,Li}}$. We are using Li as the sensor atom in the CAVS, and the dominant background gas species for most systems is $H_2$, so in many cases the *ab initio* calculations will be sufficient. In cases where other background species are present, we must use a more accurate value for $K_{\text{loss}}$, the estimates mentioned above are insufficient. As discussed in Scherschligt *et al.*,[61] we define a relative sensitivity coefficient

$$S_{GAS} \equiv \frac{K_{\text{loss}}(\text{GAS})}{K_{\text{loss}}(H_2; ab)} \cong \frac{\Gamma_{GAS}}{\Gamma_{H_2}} \qquad (13)$$

so that a careful measurement of $S_{GAS}$ combined with the *ab initio* calculations for the Li + $H_2$ system give a robust value for the loss rate coefficient of an arbitrary gas. Measuring the sensitivity coefficients of a variety of gases requires an apparatus that enables measurement of loss rates of those gases at repeatable pressures. Such an apparatus is presently under construction by the authors. The pressures do not need to be known absolutely, so a calibrated pressure gauge is not necessary, and the technique remains fundamentally primary. This apparatus is based on a traditional technique for calibrating high-vacuum gauges in which a known flow is injected from a flowmeter into a dynamic expansion chamber as described previously. To adapt this technique to our needs, ultra-low outgassing materials are used in the construction so that it can reach outgassing flux rates less than about $3\times10^{-12}$ Pa L s$^{-1}$; mostly heat-treated stainless steel[69,70] and some titanium and copper (aluminum is avoided because it reacts with alkali metals). The details of these designs will be the subject of upcoming publications.

## C. *Other loss mechanisms and error sources in the CAVS*

Now turning our attention to the issue of whether we have one-to-one correspondence between collisions and ejections, we must consider all possible ways to miscount collisions. The only realistic way we could undercount collisions is if the resulting energy transfer from the background molecule to the trapped atom is insufficient to eject it from the trap. These so-called quantum-diffractive or glancing collisions are a function of trap-depth[68] and are a small percentage of losses from a shallow trap. Because that percentage can be accurately calculated, the associated uncertainty in the pressure measurement is small. The authors are investigating this in detail and it will be the subject of upcoming publications. Ways in which one could overcount collisions are more numerous, there are a number of loss channels in the trap due to effects other than collisions with background particles. In general, trap loss is described by the following differential equation where coefficients $\Gamma$, $K_2$, and $K_3$ describe trap decay due to single-body, two-body, and three-body loss respectively (where a *body* here is a sensor atom) in a trap with density $\rho_{Li}$

$$\frac{d\rho_{Li}}{dt} = -\Gamma\rho_{Li} - K_2\rho_{Li}^2 - K_3\rho_{Li}^3. \tag{14}$$

In the event that two- and three-body losses are suppressed, the solution becomes a simple exponential as assumed in the preceding discussion. In any case, $\Gamma$, $K_2$, and $K_3$ can easily be distinguished from one another when fitting the data, and the loss rate of interest, $\Gamma$, can be extracted. Three-body loss is negligible for trapped Li at the temperatures (< 1 mK) and densities (~$10^{10}$ cm$^{-3}$) relevant for the CAVS. Two-body loss is present in the CAVS due to evaporation, where two cold atoms elastically collide exchanging energy and causing one of the atoms to be ejected from the trap. Evaporation can be controlled by raising the trap depth relative to the temperature of the sensor atoms. Models of the evaporation process are accurate and can make the associated uncertainty negligibly small.

There are several varieties of atom trap that could be used for sensing vacuum, including magnetic traps and magneto-optical traps (MOTs). MOTs use light pressure forces from lasers to trap, while magnetic traps use only magnetic fields. Our goal is to produce a primary standard as well as a sensor, so for the CAVS, we will use a pure magnetic trap to avoid complications arising from atom-laser interactions (for example, complications arising from the fraction of atoms in an excited state). In addition, the number of glancing collisions (and their associated uncertainty) is far smaller in magnetic trap (<1 %) than for a MOT (approximately 50 %), because the trap depth of a magnetic trap can be made arbitrarily small, whereas a typical MOT has a trap depth $U/k_B \sim 1$ K. Once the CAVS is established in a pure magnetic trap, it may be possible to bootstrap to other trap technologies in similar devices.

Even in a magnetic trap, there are other losses to consider. Besides collisional losses, the CAVS could potentially have Majorana losses—a trapped atom may switch from a trapped state to an un-trapped state if it passes near a region in which the magnetic field is zero.[71] This is suppressed by using a magnetic field configuration that has no field zero, though this loss channel cannot be eliminated completely and may represent non-negligible uncertainty at the lowest pressures.[72] Or the CAVS could exhibit losses due to noise in the trap leading to heating; all effort will be made to minimize noise, and to quantify the effect of what little noise remains.

While the CAVS is still under construction, we have recently operated a magneto-optical trap and used it to sense pressure (albeit in an incomplete apparatus with a configuration not optimal for testing). This MOT (the CAVS-MOT) will, in the final apparatus, load atoms into the magnetic trap used for the CAVS. In Fig. 7a, we show the decay of atom number trapped in the CAVS-MOT. At early times, we see the contribution from two-body loss (caused by light-assisted collisions) followed by exponential decay at long times. We separate the two mechanisms by fitting and extract from the exponential decay a pressure. We compare the measured pressure to that measured by an ionization gauge, as shown in in Fig. 7b. The disagreement between the measured pressures is due to the following: First, the pressure in the chamber was produced by the outgassing induced by heating a Li source, rather than in a controlled way using an injected gas. Second, a significant pressure gradient existed between the ionization gauge and the CAVS-MOT. Third, the background gas composition was unknown, and could include significant portions of $H_2$, $N_2$, $CO_2$. Fourth, the trap was not characterized well enough to determine its depth. To analyze the data, we assumed the gas was $H_2$ and took number of glancing collisions to be zero. Given the pressure gradients, we expect the ionization gauge to read lower than the CAVS-MOT, as shown in Fig. 7b. Moreover, there is excellent linear agreement between the ionization gauge and the CAVS-MOT. The preliminary data of Fig. 7b indicate that the CAVS will be a good pressure sensor. With additional effort to understand and quantify the loss mechanisms and sources of uncertainty, we will fully characterize and qualify the CAVS as a primary standard as well as an absolute sensor.

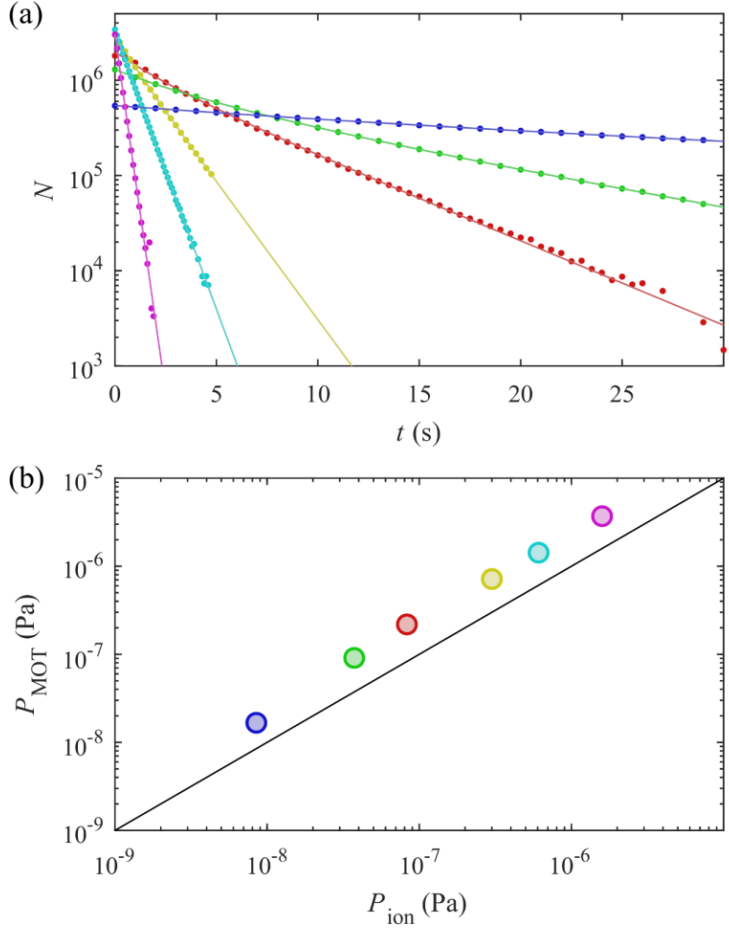

FIG. 7. (Color Online) Panel (a): Atom number decay in a magneto-optical trap (CAVS-MOT) data (circles) are fit to decay curves (solid curves) which are solutions to Eq. 14 and include single-body and two-body interactions. Panel (b): Pressure in the CAVS-MOT as determined by the data in panel (a) converted to pressure using semi-classical cross section estimates plotted versus an uncalibrated ion gauge.

## D.  Beyond the laboratory-scale CAVS

The CAVS will be the first primary pressure standard operating at UHV pressures and below at any national metrology institute. While it promises vast improvements over existing measurement technology, there remains one disappointing fact: it is confined to the lab and can only be operated by NIST personnel. By its quantum-SI nature, it is inherently accurate and never requires calibration, but it is not deployable. To truly revolutionize pressure metrology at UHV and below, we are developing a miniaturized version of the CAVS (the PICO-VS) as part of the Cold Core Technology (CCT) program.

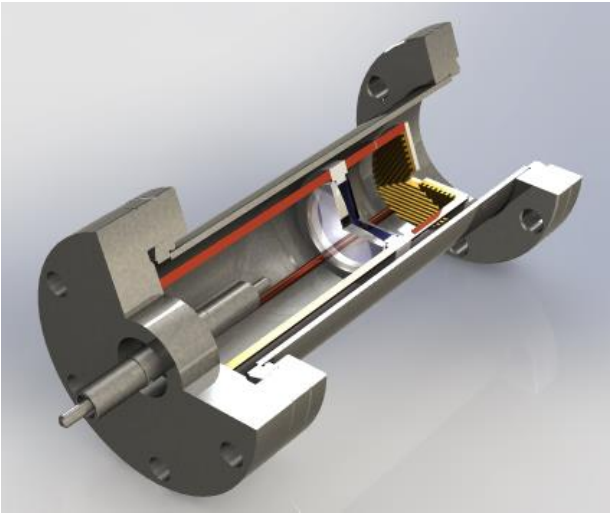

FIG. 8. (Color Online) A 3D-model of the PICO-VS, including a model of the triangular grating chip. Reprinted from S. Eckel, D. Barker, J. Fedchak, N. Klimov, E. Norrgard and J. Scherschligt, Metrologia (submitted 2018).

At its heart, CCT is a toolkit under development that enables use of cold trapped atoms for any number of applications. Like the lab-scale CAVS, the PICO-VS will use trapped Li atoms, which have several advantages over the more commonly used Rb. Rb is more easily trapped, and thus the associated technology is more mature and affordable. However, the high vapor pressure of Rb limits its use as a vacuum measurement tool for two reasons: a Rb-based device cannot be baked (and baking is essential for achieving UHV pressures) and Rb will eventually pollute the vacuum environment. Li has an exceptionally low vapor pressure ($3.2 \times 10^{-18}$ Pa at 20 °C)[73] which prevents vacuum pollution and permits Li-based devices to be baked at 150 °C.

There are several challenges to the realization of the PICO-VS. First, lithium must be heated to > 350 °C to allow production of a sufficiently large cloud of sensor atoms. Although the Li itself will rapidly stick to unheated surfaces within the sensor and not pollute the vacuum system, any contaminants outgassing from the Li source could adversely affect the vacuum pressure. Second, a MOT typically requires optical access to the Li atoms along three orthogonal axes and many optical components. The complexity and footprint of a MOT's optics will need to be reduced to make a deployable sensor. In addition to these main challenges, there are others related to production of miniaturized electromagnets, for a magnetic trap, and compact laser systems. However, these two problems have largely been solved by other research groups and by industry, so they are not a primary focus of our effort.[74,75]

We are investigating methods to reduce vacuum contamination by the PICO-VS's Li source. In air, Li oxidizes and reacts with other gas constituents to form hydroxides, nitrides, and carbonates; presumably, these compounds then contribute to outgassing and contamination when the lithium is heated. We have developed a miniature Li oven made of 3D-printed titanium.[76] This oven achieves an outgassing rate of $5(2) \times 10^{-7}$ Pa L s$^{-1}$ at operating temperature, which is approximately ten times lower than similar commercial Li sources. The outgassing rate of the oven is only limited by nitrogen contamination of the loaded Li metal and can therefore be reduced with straightforward improvements to our Li preparation. Another low-outgassing technique for producing Li vapor is light-induced atomic desorption (LIAD) which has been demonstrated for Rb and Na.[77,78] When a Li-coated surface is exposed to UV light, Li atoms desorb from the surface and can be captured by a MOT. We have loaded Li atoms from a LIAD source into a MOT in sufficient quantities for operation of the PICO-VS (although the 3D-printed titanium source allows MOT loading at a much higher rate.)[76] A LIAD source is ideal for measuring XHV pressures because it is non-thermal: any vacuum pressure increase will be rapidly erased when the UV light is extinguished. Both the 3D-printed titanium source and the LIAD source could be used to realize the PICO-VS; the preferable source will be determined by the target measurement range and necessary Li loading rate.

The optical access requirements of a MOT can be substantially reduced by using diffraction gratings. A single laser beam incident upon a 2D diffraction grating can generate all the beams needed to form a MOT. Such an optical configuration has been used to trap Rb.[79] We are currently adapting this technique for trapping of Li, which is complicated by the high operating temperature of Li sources and the comparatively weaker confinement of diffraction grating MOTs compared to traditional MOTs. A photograph of our nanofabricated grating chip is shown in Fig. (9). A Li grating MOT, combined with a low-outgassing Li source, in a suitably compact package can form a first generation of the PICO-VS. The lowest detectable pressure for this device will likely be limited by the large depth of the MOT (see section V.C.) and will be the subject of future study. The second generation PICO-VS (the PICO-VS2) will integrate a miniaturized magnetic trap to allow primary sensing of even lower pressures. We have recently demonstrated a grating based Li MOT which will be the subject of upcoming publications.

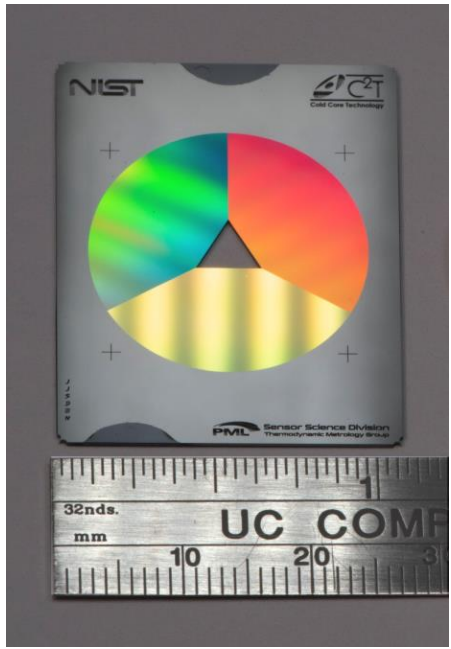

FIG. 9. (Color Online) Photograph of the prototype CCT triangular grating chip, with ruler.

## E. Uncertainty and the pressure limits

A large part of NIST's research in the CAVS is understanding the sources of uncertainty, as discussed in sections V.b – V.d. We anticipate that the uncertainty will be dominated by the collision cross-section calculation; the other sources of uncertainty discussed above are anticipated to be small by comparison. As we discussed, the calculated cross-sections should have an uncertainty below 5 % ($k=2$) when completed.

In use, there is an additional uncertainty due to gas composition. If a process or calibration gas is leaked into the vacuum system under test to a pressure of at least a factor of three above the background, then this uncertainty will be insignificant. However, when using the CAVS to determine the background pressure, this uncertainty must be considered. This is system dependent, but the majority of vacuum systems will have background pressures dominated by the partial pressure of $H_2$. Because the variation in loss-rate for common constituents (other than $H_2$) of the background gas is expected to be only about 20 % from semi-classical estimates, even limited knowledge of the background gas composition should, in most cases, give uncertainties on the 20% level.

The collisional time-scale for atom loss in the trap varies by roughly ($2 \times 10^{-6}$ Pa)/p. At $1 \times 10^{-6}$ Pa this means the expected trap lifetime in roughly 0.2 seconds. Faster lifetimes than this are practically very difficult to determine because of difficulties of loading a trap with sensor atoms. For this reason, cold-atom experiments are typically performed deeper in the UHV. The lowest pressure that can be absolutely measured by the CAVS will be limited by low-frequency noise, magnetic field noise leading to trap heating, or loss due to non-adiabatic spin flips, discussed above. This is under active research by our group at NIST and is unknown at this time. With careful design, these factors are not likely to limit the CAVS for pressures above $1 \times 10^{-9}$ Pa, and it may be that much lower pressures are possible, likely extending into the XHV but with somewhat higher uncertainty.

## VI. SiN RING-DOWN MEMBRANE GAUGE (THE "BRANE" GAUGE)

Mechanical damping by drag forces on many types of structures has been used for a wide range of vacuum pressure sensors.[80] Broadly two classes of devices exist, levitated spinning rotors (or spinning rotor gauges)[81], which notably have been employed as stable transfer standards for high vacuum[82] and oscillating mechanical resonators, ranging from MEMs devices to macroscopic pendula and torsional oscillators. Generally, these gauges are desirable because they act as absolute pressure sensors with high linearity, operate at high frequencies away from DC to minimize low frequency noise and drift, often allow for direct computation of pressure dependence from first principles, and do not generate large amounts of heat. The linear dynamic range is limited by intrinsic mechanical dissipation at low pressure, and the transition from molecular flow to viscous damping at high pressure. So the ideal properties of such gauges include low intrinsic mechanical dissipation and a mechanical element that is smaller than the mean free path of the gas being measured at the highest pressures of interest. A larger mechanical element would still exhibit pressure-dependent damping, but it would not be characteristically linear.[83–85]

We are interested in developing a mechanical damping gauge for high vacuum in range of $10^{-6}$ to $10^{-2}$ Pa, which is simple, robust, and sufficiently stable to operate as a sensor and transfer standard. The successful combination of this device with the FLOC and the CAVS will constitute a suite of instruments that covers the entire pressure range from a few atmospheres to the lowest achievable laboratory vacuums. Recent experimental progress in the field of quantum optomechanics has led to the development of optically detected and actuated mechanical resonators, well suited for gas damping pressure sensing. The mechanical sensing element consists of an ultralow intrinsic damping rate, $\Gamma_i$, silicon nitride membrane, whose out-of-plane drumhead modes (Fig. 10) have ultrahigh intrinsic mechanical quality factor ($Q = \omega_m/\Gamma_i$), approaching 1 billion. With millimeter scale transverse extent and thickness, $h$, less than 30 nm, these mechanical resonators are readily damped by surrounding gas, while retaining resonance frequencies, $\omega_m/2\pi$, in the hundreds of kilohertz. These devices allow for simple optical readout, are insensitive to low frequency vibration, magnetic fields, and sensor tilt, and do not require active stabilization. The total mechanical damping rate, the inverse of the mechanical ringdown time, $\tau_{rd}$, in the molecular flow regime is given by[83]

$$\frac{1}{\tau_{rd}} = \Gamma_{tot} = \Gamma_i + p\sqrt{\frac{32}{\pi}\frac{m_m}{k_B T}}\,\Big/\rho h \;, \tag{15}$$

where $m_m$ is the molecular mass of the gas at pressure $p$, and $\rho$ is the density of silicon nitride. Recent devices have demonstrated that sub-mHz intrinsic damping rates are achievable,[86–88] equivalent to the damping from air pressure in the $10^{-5}$ Pa range. For such devices, we estimate the transition region to the viscous flow regime lies above 1 Pa, implying large dynamic range gauges should be possible.

Figure (10) shows preliminary results for a silicon nitride membrane mechanical damping gauge. We mechanically excite the membrane with a piezoelectric actuator and measure the energy ringdown time with a simple optical interferometer. We demonstrate a linear dynamic range of over two orders of magnitude, limited by excess dissipation of mechanical energy into the membrane mounting structure. Devices with optimized geometry and mounting should extent the dynamic range by several more orders of magnitude,[86–88] as well by employing higher order mechanical modes of the membrane.[85,86] We find the slope sensitivity in the linear region of Fig. (10) agrees with the prediction of Eq. (15) at the approximately 10 % level, limited by our uncertainty in the membrane thickness and density.

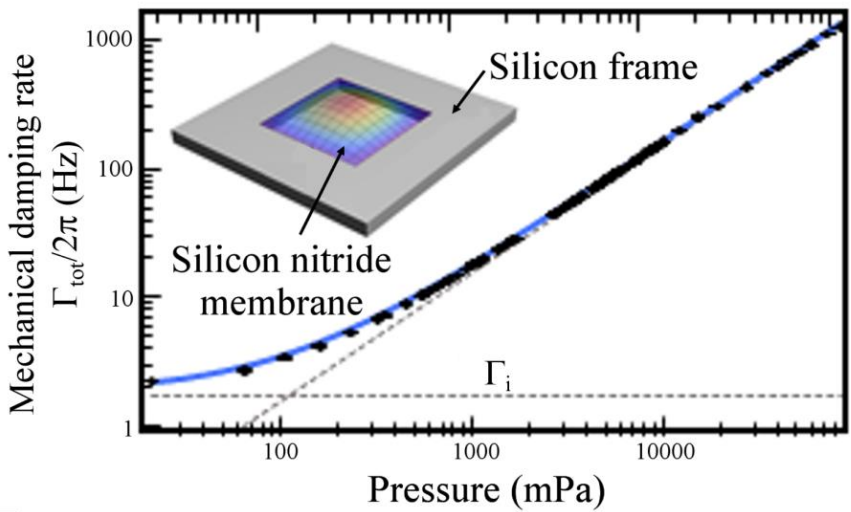

FIG. 10. (Color Online) Silicon nitride membrane mechanical damping gauge. The ringdown time for a 2 mm square by 50 nm thick, high-tensile-stress membrane is measure via piezoelectric actuation and optical detection. Inset shows fundamental out-of-plane vibrational mode.

## VII. SPECTROSCOPIC TECHNIQUE FOR MEASUREMENT OF TRANSIENT PRESSURE

As discussed throughout this manuscript, we are pursuing several methods to accurately measure static pressure from XHV to tens of MPa. However, there are no commercially available traceable calibration methods for the measurement of transient pressure. The measurement of transient pressure is important for many applications, perhaps chief among them is to understand the effect on the human brain of explosions or rapid accelerations such as in an automobile crash, which the authors expect to lead to better safety standards and equipment design. In our method, we rely on the traceability method outlined above, the unique quantum mechanical characteristics of the molecules are themselves the standard for pressure, making it consistent with the quantum-SI.

The goal of recent European NMI's via EURAMET EMRP IND09 is to achieve traceable measurement of transient pressure using quantitative modeling of shock tube dynamics.[89,90] In contrast, our approach is to use independent molecular spectroscopy as a dynamic measurement of pressure, where the pressure itself is ascertained by measuring time-resolved pressure-broadened spectra of CO molecules.[91,92] For our application, the shock tube is only used to produce a step change in pressure, i.e. act as a transient pressure source. From the linewidth and intensities of ro-vibrational transitions, pressure and temperature can be determined. For transient pressure measurement, our goal is to achieve an absolute uncertainty of 5 % with a measurement rate of 100 kHz.

We have recently constructed and characterized a dual diaphragm shock tube that allows us to achieve shock amplitude reproducibility of approximately 2.3 % for shocks with Mach speeds ranging from 1.26 to 1.5.[93] The agreement to 1-D modeling over this limited range is within a few percent and we believe a limiting factor in assessing the 1-D model is the inherent limitation of the piezo electric sensors used to determine Mach speed of the propagating shockwave. The large area sensors have spatial averaging effects which limit the accuracy in determining the time of the shock. Additionally, acceleration effects, temperature dependence, low resonant frequency, and over/under-shoot in these devices dominate the noise as one moves to high amplitude shocks. Figure (11) illustrates the piezo electric sensors response to a shock wave traveling at Mach 1.8. To overcome these challenges, we are developing phonic sensors that have extremely fast rise times (ns) and very small sensing area (100 μm).

In a proof-of-concept study we used our shock tube to characterize the dynamic response of photonic sensors embedded in polydimethylsiloxane (PDMS), a material of choice for soft tissue phantoms. Our results indicate that the PDMS-embedded photonic sensors response to shock evolves over tens to hundreds of microseconds time scale making it a useful system for studying transient pressures in soft tissue.

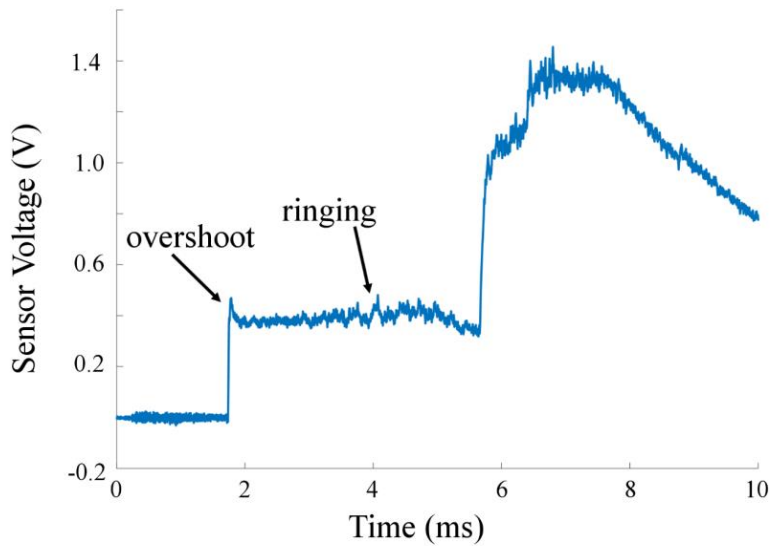

FIG. 11. (Color Online) The (blue) trace was recorded from a piezo electric transducer mounted perpendicular to the shock front. The Mach speed of the shock was measured to be 1.8. The initial conditions were 1.9 MPa and atmospheric pressure using Nitrogen.

## VIII. SUMMARY AND CONCLUSIONS

Quantum-based devices have great promise for metrology. This is readily seen in time and frequency metrology, the realization and dissemination of the second is now an entirely quantum-SI enterprise. Adoption of this new paradigm happened swiftly with wide acceptance, so much so that even a teenager checking the time on her phone becomes an unwitting quantum-SI metrologist. The advantage of a quantum-based standard is that it's always correct (within its uncertainty and unless it's broken) because of the invariance of the physical constants and laws upon which it depends—the charge on one electron is the same as the charge on any electron, and so the charge on ten million electrons is exactly ten million times the charge on one electron. The uncertainty comes from the errors in counting to ten million, which may come from instrumentation, calculations, and noise. The practical benefit of transitioning to this new paradigm is that a quantum-SI device never needs to be calibrated, so saving enormous cost and effort, and is said to have a zero-length traceability chain. This paper focused on NIST's current efforts to recast the pascal in quantum terms, but we can expect more projects to come online that will enable us to cover the entire range from the pressures of deep interstellar space to the pressures in an explosion.

## ACKNOWLEDGMENTS

We thank Václav Havel, and the contributions of a hopeful person.